\documentclass[10pt]{IEEEtran}

\usepackage{float}
\usepackage{subfigure}
\usepackage{stfloats}
\usepackage{cite}
\usepackage{amsmath,amssymb,amsfonts}
\usepackage{graphicx}
\usepackage{textcomp}
\usepackage{mathrsfs}
\usepackage{cases} 
\usepackage{algorithm}
\usepackage{algorithmicx}
\usepackage{algpseudocode}
\usepackage{color}
\usepackage{amsmath}

\usepackage{comment}
\usepackage{titlesec}

\usepackage{tabularx}
\usepackage{booktabs}

\titlespacing{\subsection}{0pt}{5pt}{5pt}
\titlespacing{\section}{0pt}{5pt}{5pt}

\newcommand{\be}{\begin{equation}}
\newcommand{\ee}{\end{equation}}
\def\bee#1\eee{\begin{align}#1\end{align}}
\newcommand{\bse}{\begin{subequations}}
\newcommand{\ese}{\end{subequations}}

\ifCLASSINFOpdf
\else
\fi

\begin{document}
	
	\title{Technical Report: A Hierarchical Dynamically Weighting Deep Reinforcement Learning Method for Multi-UAV Multi-Task Coordination}

	\author{Xindi Wang, Haining Li, Tao Ding, Bolin Cai
	\thanks{All authors are with the School of Internet, Anhui University, China; Corresponding author: Bolin Cai (email: cbl@ahu.edu.cn). 
		
	This work was supported in part by NSFC under Grant 62202007.}
}
\maketitle

\begin{table*}[t]
	\caption{Notations Table}
	\label{tab:notation}
	\centering
	\small
	\renewcommand{\arraystretch}{1.15}
	
	\begin{tabular}{p{0.18\textwidth} p{0.72\textwidth}}
		\toprule
		\textbf{Notation} & \textbf{Meaning} \\
		\midrule
		
		$M$ & The set of UAVs deployed in the scenario. \\
		
		$N$ & The set of GUs deployed in the scenario. \\
		
		$v_t$ & The velocity of the user at time slot $t$. \\
		
		$\theta_t$ & The heading angle of the user at time slot $t$. \\
		
		$\alpha$ & Memory factor of the Gauss-Markov mobility model. \\
		
		$w_{v,t}$ & Zero-mean Gaussian noise term for the velocity process at time slot $t$. \\
		
		$w_{\theta,t}$ & Zero-mean Gaussian noise term for the heading angle process at time slot $t$. \\
		
		$h$ & The ﬁxed ﬂight altitude of the UAV. \\
		
		$C^t$ & The image acquisition state matrix. \\
		
		$h_{m,n}(t)$ & The channel coefficient between GU $n$ and UAV $m$ at time slot $t$. \\
		
		$\alpha_{m,n}$ & The channel power gain at the reference distance of $\left\| L_m^t - L_n^t \right\| = 1\,\mathrm{m}$. \\
		
		$B$ & The total system bandwidth. \\
		
		$R_{m,n}(t)$ & The achievable communication rate between UAV $m$ and GU $n$ at time slot $t$. \\
		
		$p_{tx}$ & The transmit power of UAV. \\
		
		$D_n(t)$ & The data queue length of GU $n$ at time slot $t$. \\
		
		$p_{m}^{prop}(t)$ & The propulsion power of UAV $m$ at time slot $t$. \\
		
		$\bar{p}$ & The energy budget of each UAV. \\
		
		$\gamma$ & The discount factor. \\
		
		$w_{cov}$ & The weighting coefficients associated with image acquisition. \\
		
		$w_{comm}$ & The weighting coefficients associated with communication. \\
		
		$\mathcal{C}(T)$ & The completion ratio of image acquisition. \\
		
		$\mathcal{R}(T)$ & The completion ratio of communication. \\
		
		$\underline{h}$ & The channel coefficient threshold for satisfying the QoS requirement. \\
		
		$Q(s_t, a_t)$ & The action-value functions. \\
		
		$\alpha_{\mathrm{ema}}$ & The exponential moving average smoothing coefficient. \\

		$x_t$ & The local state of UAV at time slot $t$. \\

		$w_k^{ep}$ & The episode-level weight. \\

		$w_{t}^{st}$ & The step-level weight. \\

		$w_t$ & The fused weight of the episode-level weight and the step-level weight. \\

		$s_k^{ep}$ & The global state at the $k$-th episode \\

		$\bar{C}_{k-1}$ & The exponential moving averages of the image acquisition completion rate at $(k-1)$-th episode. \\

		$\bar{R}_{k-1}$ & The exponential moving averages of the communication completion rate at $(k-1)$-th episode. \\

		$\alpha_{k-1}$ & The weight coefficient learned in the $(k-1)$-th episode. \\

		$P_m^t$ & The coordinates of UAV $m$. \\

		$U_m^t$ & The coordinates of users performing communication uploads to UAV $m$. \\

		$C_m^t$ & The local image acquisition state matrix centered on the current UAV $m$. \\

		$g_t$ & The global context state. \\
		
		$\tau$ &  Softmax temperature in step-level weight net. \\

		$\varepsilon$ & The greedy coeﬃcient in action selection of the UAV. \\

		$\delta_t$ & The hierarchical weight-mixing coefficient. \\

		\bottomrule
	\end{tabular}
\end{table*}

\begin{abstract}
This paper investigates the multi-UAV multi-task coordination problem in infrastructure-less emergency scenarios, where UAVs collaboratively are required to jointly perform aerial image acquisition and ground-user communication. To tackle the challenge of balancing heterogeneous tasks within dynamic environments, we propose a hierarchical dynamic weighting Deep Reinforcement Learning (DRL) framework. Specifically, an episode-level module is introduced to capture global task preferences, while a step-level module adaptively adjusts the objective weights according to real-time system conditions. By integrating global and instantaneous weights, the proposed framework improves decision stability and responsiveness during task execution. Simulation results demonstrate that the proposed method achieves faster convergence, more stable training, and higher task completion efficiency than conventional works.
\end{abstract}

\begin{IEEEkeywords}
unmanned aerial vehicles, multi-task coordination, deep reinforcement learning, dynamic weighting.
\end{IEEEkeywords}

\IEEEpeerreviewmaketitle

\section{Introduction}
With the development of cooperative multi-UAV systems for single-task execution, research interest has increasingly shifted toward intelligent multi-UAV multi-task systems. In such systems, multiple UAVs are required to collaboratively accomplish heterogeneous tasks under spatiotemporal constraints and limited onboard resources\cite{lei2022multitask,wang2025hierarchical}. However, due to the strong interdependence among subtasks and the potential conflicts between objectives, it remains challenging to simultaneously achieve all task requirements in dynamic environments\cite{sun2024multi}.

Recently, Deep Reinforcement Learning (DRL) has demonstrated great potential in solving long-term sequential  decision-making problems in dynamic environments\cite{wang2026multi}. 
Nevertheless, in DRL-based multi-UAV multi-task systems, the optimization of multi-UAV decision-making relies primarily on environmental reward feedback for their actions, which remains one of the core issues in current system design 
\cite{lv2023multi}.
In general, the overall reward of multi-task is typically formulated as a weighted aggregation of the rewards of individual subtasks. In the context, some works design the weights in a static way, relying on manual tuning \cite{Li2024Robust,Pan2025IoT} or static pre-configuration\cite{mu2025preference,yu2021multi}, resulting in an inability to dynamically adjust to system features, and thus affecting overall performance.
To overcome the limitations of static weighting schemes, recent works have explored dynamic reward weighting mechanisms \cite{song2024energy,gao2023mo,huang2024multiobjective} to improve adaptability to environmental variations. However, the existing methods generate weights mainly based on current scene and historical information, which causes significant fluctuations in the reward function. As a result, this reward inconsistency, induced by time-varying preferences, can bias value estimation and degrade policy learning performance\cite{huang2024multiobjective}.
In summary, in the considered multi-UAV multi-task dynamic systems, due to the drawbacks of weight design, existing DRL frameworks fail to efficiently adapt to dynamic environments where the importance of different subtasks may vary over time as resource availability and task completion status evolve. Consequently, reward design remains a major bottleneck in current DRL-based solutions.

To address these challenge, a \textbf{H}ierarchical \textbf{D}ynamically \textbf{W}eighting mechanism based multi-agent multi-task \textbf{DRL} (\textbf{HDWDRL}) framework  is proposed here, in which an episode-level Actor-Critic module is introduced at the weight-learning layer to capture global prior weight distributions, and a step-wise state-aware weighting network is used to analyze real-time system conditions. By combining the two above modules to construct a dynamic weighting framework, enabling balancing balance objectives among subtasks and adapt to the current environmental conditions. Simulation results demonstrate that, under the considered scenarios, the proposed framework exhibits better adaptability to dynamic environments, maintains stable training behavior, and improves execution efficiency in multi-task scenarios compared with conventional methods.

\section{System Model and Problem Formulation}
\subsection{System Model}
This paper considers an emergency scenario where terrestrial communication infrastructure is completely unavailable. In such a situation, a multi-UAV system is deployed to simultaneously accomplish two heterogeneous tasks. The first task is aerial image acquisition for rapid disaster reconnaissance. The second task is to provide reliable emergency data links for densely distributed ground users (GUs). To support real-time coordination in the multi-UAV multi-task system, each UAV is equipped with an independent decision-making module, which determines its actions based on local observations and neighboring information exchanged through communication links. Moreover, UAVs make decisions sequentially according to their assigned priorities within each time slot. Through this information-sharing mechanism, UAVs can achieve conflict-free mission allocation in a decentralized and efficient manner.\footnote{For clarity, the main notations used throughout this paper are summarized in Table.~\ref{tab:notation}.}

In the aforementioned application scenarios where a set of UAVs ($M = \{1, \dots, m\}$) are deployed and a set of GUs ($N = \{1, \dots, n\}$.) are spatially distributed and move with velocity $v_t$ and heading $\theta_t$ \cite{sheng2025outage} as :
\begin{align}
	v_t &= \alpha v_{t-1} + (1-\alpha)\overline{v} + w_{v,t},  \\
	\theta_t &= \alpha \theta_{t-1} + (1-\alpha)\overline{\theta} + w_{\theta,t},
\end{align}
where $w_{v,t} \sim \mathcal{N}\bigl(0, \sigma_v^2\bigr), w_{\theta,t} \sim \mathcal{N}\bigl(0, \sigma_\theta^2\bigr)$. The target area is discretized into two-dimensional grids with size $H \times W$ \cite{rizvi2024multi}.

$\blacksquare$ \textbf{Image Capture Model}: During flight, UAVs capture ground images via onboard cameras and extract useful information. The horizontal and vertical fields of view are given by $H_{FOV} = 2h \tan\left(\frac{\varphi_h}{2}\right)$ and $V_{FOV} = 2h \tan\left(\frac{\varphi_v}{2}\right)$ where $h$ denotes the UAV altitude, and $\varphi_h$ and $\varphi_v$ are the horizontal and vertical viewing angles, respectively. The UAVs take photos as they fly to the center of the grids, in which the image acquisition state matrix $C^t\in\mathbb{R}^{H \times W}$ at time slot $t$ is given by
\begin{align}
	C^t[i,j] =
	\begin{cases}
		1, & \text{if the cell } (i,j) \text{ has been captured,} \\
		0, & \text{otherwise.}
	\end{cases}
\end{align}

$\blacksquare$ \textbf{Communication Model}: The GU-to-UAV wireless channel is modeled as a Rician fading channel \cite{wang2023trajectory}. The channel coefficient between GU $n$ and UAV $m$ at time slot $t$ is expressed as
\begin{align}
	h_{m,n}(t) = \sqrt{\beta_{m,n}(t)} \, \tilde{h}_{m,n}(t),
\end{align}
where $\beta_{m,n}(t) = \frac{\alpha_{m,n}}{\left(h^2 + \left\| L_m^t - L_n^t \right\|^2\right)^{K_{ps}/2}}$ and $\tilde{h}_{m,n}(t) = \sqrt{\frac{K_s}{K_s+1}} \tilde{h}_L + \sqrt{\frac{1}{K_s+1}} \tilde{h}_N \ $ with $|\tilde{h}_L| = 1$  and $\tilde{h}_N \sim \mathcal{CN}(0,1)$. Here, $K_{ps}$ and $K_s$ denote the path loss exponent and the Rician factor, respectively. The parameter $\alpha_{m,n}$ represents the channel power gain at a reference distance of $\left\| L_m^t - L_n^t \right\|=1\mathrm{m}$.
By adopting OFDMA technology \cite{wang2023trajectory}, the UAV pre-divides the available communication resources into multiple equal and orthogonal resource blocks, and serves at most $N$ GUs simultaneously. Given a total system bandwidth $B$ equally partitioned into $N$ sub-channels, the bandwidth allocated to each sub-channel is $B_s = \frac{B}{N}$.
Accordingly, the achievable communication rate between UAV $m$ and GU $n$ at time slot $t$ is expressed as
\begin{equation}
	R_{m,n}(t) = B_s \log_2 \left( 1 + \frac{|h_{m,n}(t)|^2 p_{tx}}{\sigma^2} \right),
	\label{eq:rate}
\end{equation}
where $|h_{m,n}(t)|^2$ denotes the channel power gain, $p_{tx}$ is the transmit power of UAV $m$, and $\sigma^2$ represents the additive white Gaussian noise (AWGN) power. Meanwhile, the data queue length of GU $n$ at time slot $t$ evolves as $D_n(t) = D_n(t-1) - R_{m,n}(t)\, t_{\text{comm}}$ where $t_{\text{comm}}$ denotes the communication transmission duration of GU $n$ within a single time slot.

\subsection{Problem Formulation}
Under the considered scenario, the objective is to derive a policy for the multi-UAV multi-task system that minimizes the total time $T$ required to complete the tasks, formulated as
\begin{align}
	\textup{P1}: \pi^* &= \arg \min_{\pi \in \Pi} T \label{eq:problem} \nonumber\\
	\text{s.t.} \quad
	\sum_{t=1}^{T} p_{m}^{prop}(t) &\leq \bar{p}, \ \forall m \in M \tag{a} \\
	\mathcal{C}(T) &\geq \rho_c,\ \mathcal{R}(T) \geq \rho_r, \tag{b} \\
	h_i^t &\geq \underline{h}, \tag{c} \\
	P_i(t) &\neq P_j(t), \forall i, j \in M,\ i \neq j \tag{d}, 
\end{align}
where $\pi^*$ denotes the optimal policy and $\Pi$ is the feasible policy space. In (a), $p_m^{\mathrm{prop}}(t)$ denotes the propulsion power of UAV $m$ at time slot $t$, and $\bar{p}$ is the energy budget of each UAV. Constraint (b) requires that the completion ratio of image acquisition satisfies threshold $\mathcal{C}(T) \geq \rho_c$, where $\mathcal{C}(t) = \frac{1}{H \cdot W} \sum C^t[i,\,j]$, and that the completion ratio of communication satisfies $\mathcal{R}(T) \geq \rho_r$, where $\mathcal{R}(T)=1-\frac{\sum_{n=1}^{N} D_n(T)}{\sum_{n=1}^{N} D_n(0)}$. 
Constraint (c) ensures the QoS of communication channel, and constraint (d) prevents collisions among UAVs.

\section{Multi-Agent Multi-Task DRL Algorithm}
Due to the fact that the optimization objective in P$1$ is intractable to solve directly, we reformulate it as a long-term reward maximization problem over $T$ time slots under the same constraints. Specifically, the original objective of minimizing task completion time is transformed into maximizing the cumulative system performance, which jointly reflects the completion rates of image acquisition and communication. The resulting optimization problem can be expressed as
\begin{align}\label{eq:reformulated_objective}	
	\pi^* = \arg \max_{\pi} \mathbb{E}\left[ \sum_{i=0}^{T-t} \gamma^i \left( w_{cov} \mathcal{C}(t+i) + w_{comm} \mathcal{R}(t+i) \right) \right]		
\end{align}
where $\gamma \in (0,1]$ is the discount factor, and $w_{cov}$ and $w_{comm}$ denote the weighting coefficients associated with image acquisition $\mathcal{C}(t)$ and communication $\mathcal{R}(t)$, respectively. To solve this problem efficiently, we develop a DRL-based framework that jointly models long-term reward estimation and dynamic multi-objective weighting. The proposed framework consists of two main components: a multi-head Q-network for task-specific value estimation, and a hierarchical dynamic weighting mechanism for adaptive objective balancing. The former captures the long-term impact of UAV actions under the current policy, while the latter adjusts the relative importance of different tasks according to both episode-level and step-level system states.
\begin{figure*}[htbp] 
	\centering
	\includegraphics[width=0.92\textwidth, keepaspectratio]{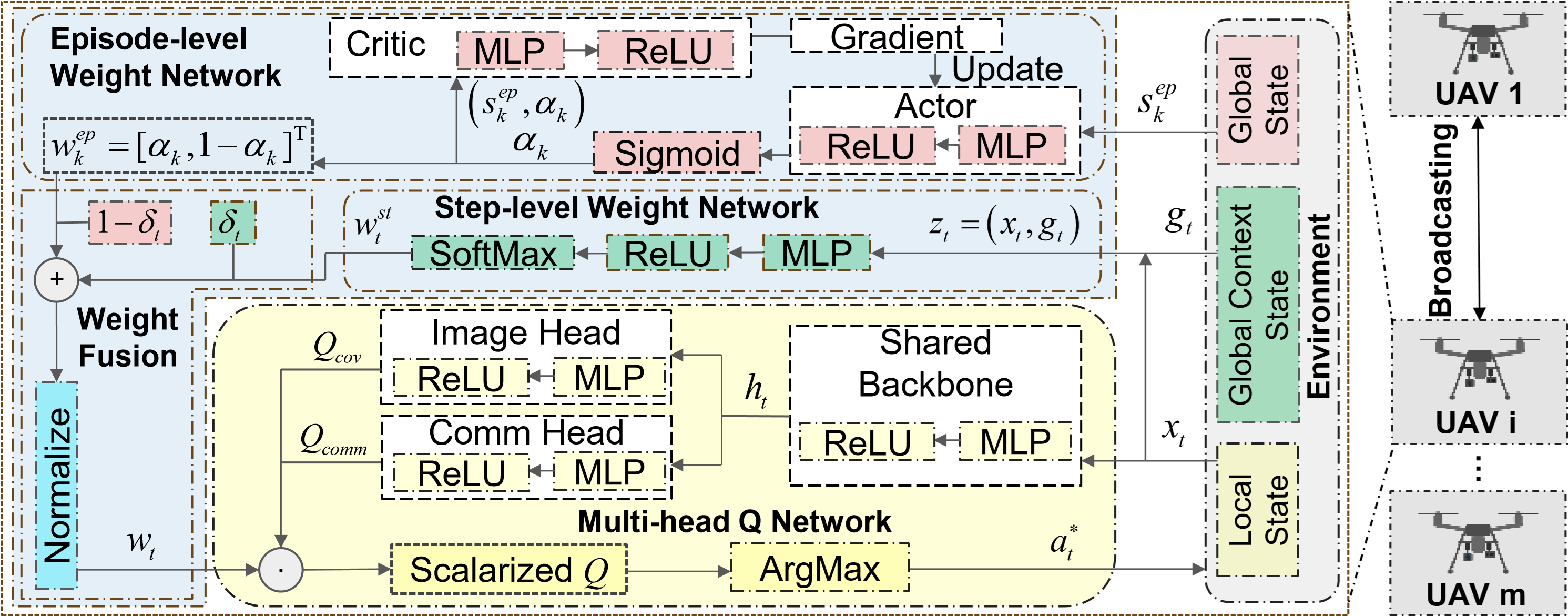}
	\caption{An illustration of the proposed HDWDRL framework}
	\vspace{-0.6cm} 
	\label{fig:1}
\end{figure*}

\subsection{DRL Framework}
To optimize the objective in (7), we design a DRL framework that decomposes the learning process into two coupled parts. The first part, shown in the yellow block of Fig.~1, models the accumulated discounted returns of the system, i.e., $\sum_{i=0}^{T-t}\gamma^i (\mathcal{C}(t+i) + \mathcal{R}(t+i)$, which captures the long-term effect of UAV actions on task completion. The second part, shown in the blue blocks of Fig.~1, learns dynamic weighting coefficients $w_{cov}$ and $w_{comm}$,  enabling the policy to adapt to changing task priorities in real time.

$\blacksquare$ \textbf{The first part}: A multi-head Q-network is employed to estimate the long-term cumulative return associated with each action under the current policy $\pi$. However, to reduce gradient interference caused by shared parameters across tasks, we adopt a multi-head architecture with a shared backbone and task-specific output heads. Accordingly, the task-specific action-value functions are defined as
\begin{align}
	Q_{cov}(s_t, a_t) &= \mathbb{E}\left[ \sum_{i=0}^{T-t} \gamma^i \mathcal{C}(t+i) |s_t, a_t \right], \label{eq:Q_cov} \\
	Q_{comm}(s_t, a_t) &= \mathbb{E}\left[ \sum_{i=0}^{T-t} \gamma^i \mathcal{R}(t+i)|s_t, a_t \right]. \label{eq:Q_comm}
\end{align}

In the proposed multi-head Q-network, the local state of UAV $m$ at time slot $t$, denoted as $x_t$, is first encoded by a shared multilayer perceptron (MLP) backbone into a latent feature representation $h_t$. The resulting feature is then fed into multiple independent value heads, each consisting of task-specific fully connected layers that map $h_t$ to the corresponding action-value estimate. In this work, two value heads are used to evaluate the image acquisition and communication objectives, respectively. The two branches are trained independently using task-specific Bellman updates, thereby decoupling value estimation and facilitating multi-task weight fusion.

$\blacksquare$ \textbf{The second part}: After obtaining $Q_{cov}(s_t, a_t)$ and $Q_{comm}(s_t, a_t)$, the remaining challenge is how to guide decision-making when the relative importance of subtasks varies across different stages in a dynamic environment. To address this issue, we propose a hierarchical dynamic weighting mechanism to model $w_{cov}$ and $w_{comm}$. The mechanism decomposes weight learning into an episode-level network and a step-level network, as illustrated in the blue block of Fig.~1, which avoids the instability and policy drift that may arise when weights are adjusted solely based on instantaneous observations.
The episode-level weight captures the global importance of objectives over an entire episode, thereby establishing long-term preference and resource allocation. In contrast, the step-level weight adapts the objective importance at each time step, allowing the policy to respond to short-term state variations. By combining these two levels of weighting, the proposed framework improves both policy stability and environmental responsiveness.

The episode-level weight $w_k^{ep}$ and the step-level weight $w_{t}^{st}$ are fused as $w_t = \text{Normalize}((1-\delta_t) w_k^{ep} + \delta_t w_t^{st})$, where $\delta_t$ is designed according to the imbalance between the two subtasks.
\subsection{Input Layer Architecture}
Based on the proposed DRL framework, the network inputs are organized into three parts. Specifically, the local state is fed into the multi-head Q-network and, together with the global context state, serves as the input to the step-level weight network. The global state is used as the input to the episode-level weight network.

$\blacksquare$ \textbf{Global State:} To capture long-term performance trends and inter-task discrepancies for episode-level preference learning, the global state at the $k$-th episode  is defined as: $s_k^{ep} = \left[ \bar{C}_{k-1}, \, \bar{R}_{k-1}, \, \zeta_k, \, \alpha_{k-1} \right]^{T}$,  where $\bar{C}_{k-1}$ and $\bar{R}_{k-1}$ denote the exponential moving averages (EMA \cite{liu2019distributed}) of the image acquisition completion rate and communication completion rate in the previous $(k-1)$-th episode, respectively. \footnote{The EMA, or its equivalent forms such as soft target updates, is widely adopted in DRL to stabilize training. It is defined as $\bar{C}_k = (1-\alpha_{\mathrm{ema}})\bar{C}_{k-1} + \alpha_{\mathrm{ema}} C_k$ and $\bar{R}_k = (1-\alpha_{\mathrm{ema}})\bar{R}_{k-1} + \alpha_{\mathrm{ema}} R_k$, where $\alpha_{\mathrm{ema}}$ denotes the smoothing coefficient.}.
$\zeta_k$ represents the normalized discrepancy between the two objectives, defined as $\zeta_k = \frac{\bar{C}_{k-1} - \bar{R}_{k-1} + 1}{2}$. The term $\alpha_{k-1}$ denotes the weight coefficient learned in the $(k-1)$-th episode. These components are concatenated in a fixed order to form a one-dimensional feature vector, which is then used as an input to the episode-level weight network.
The network takes the concatenated pair $(s_k^{ep}, \alpha_k)$ as input and outputs its corresponding value estimate.

$\blacksquare$ \textbf{Local State:} To support step-level decision-making, the local state incorporates the UAV’s own environment, its spatial relationships with neighboring UAVs, and its service associations with GUs. Specifically, it includes the coordinates of the current UAV, the coordinates of GUs uploading data to this UAV, the coordinates of other UAVs, the coordinates of GUs served by other UAVs, and the local image acquisition state matrix centered on the current UAV. For UAV $m$, the input is defined as $x_t=[P_m^t, U_m^t, P_{-m}^t, U_{-m}^t, C_m^t]^T$ where $P^t_m$ denotes the position of the current UAV $m$, $U_m^t$ denotes the state of its associated GUs, $P_{-m}^t$ and $U_{-m}^t$ represent the states of the other UAVs and their associated GUs, and $C_m^t$ is the local $3 \times 3$ image acquisition state matrix.

$\blacksquare$ \textbf{Global Context State:} To bridge the episode-level preference and the real-time task progression, we define the global context state as $g_t = [\mathcal{C}(t), \mathcal{R}(t), 1-\mathcal{C}(t), 1-\mathcal{R}(t), \alpha_k, 1-\alpha_k, \mathcal{C}(t)-\mathcal{R}(t)]^T$. This context vector encodes the completion rates of image acquisition and communication, their corresponding deficiencies, and the episode-level weight. The concatenated input $[x_t^T , g_t^T]^T$ is then fed into the step-level weight network.

\subsection{Hierarchical Dynamic Weight Networks}\label{subsect:3C}

$\blacksquare$ \textbf{Episode-level Weight Network:} 
As shown in the blue box at the top of Fig.~1, to provide a stable global baseline by capturing long-term task progression and mitigating oscillations caused by transient step-level variations, the global state $s_k^{ep}$  is fed into the Actor network $f_\phi(\cdot)$ which is implemented as an MLP with ReLU activations in the hidden layers and a sigmoid output layer to produce a normalized scalar weight.
The episode-level weight is obtained as $\alpha_k = f_\phi(s_k^{ep})$, and the corresponding weight vector is defined as $w_k^{ep} = [\alpha_k, \, 1-\alpha_k]^T$. 
The Critic network then evaluates the state-weight pair $(s_k^{ep}, \alpha_k)$ and outputs the value estimate $V_\psi(s_k^{ep}, \alpha_k)$, which is used to guide network updates. 

$\blacksquare$ \textbf{Step-level Weight Network:} 
As shown in the blue box in the center of Fig.~1, to enable the network adaptively reflect the current task status while preserving consistency with the episode-level preference, the designed step-level weight network $f_\xi(\cdot)$ takes the local observation $x_t$ and the global context $g_t$ as input at each decision step $t$. The concatenated input vector $[x_t^T, g_t^T]^T$ is processed by two fully connected layers with ReLU activations, and the network outputs a normalized step-level weight via a temperature-scaled softmax function $w_t^{st} = \text{Softmax}(f_\xi([x_t^T, g_t^T]^T)/\tau)$, where $\tau$ controls the sharpness of the output distribution. Then, the step-level weight network is trained online using a self-supervised scheme. Specifically, a heuristic target weight is constructed as $w_t^{target} = \text{Normalize}(\lambda_1 G_t + \lambda_2 b_t + \lambda_3 w_k^{ep})$ where $G_t = \text{Normalize}([\max(r_{cov},0), \max(r_{comm},0)])$ represents immediate positive rewards and $b_t = \text{Normalize}([1-\mathcal{C}(t), 1-\mathcal{R}(t)])$ captures task completion deficiencies. The coefficients $\lambda_1$, $\lambda_2$, and $\lambda_3$ determine the relative contributions of these terms. The step-level weight network is then updated by minimizing $L_\xi^{st} = \| w_t^{st} - w_t^{target} \|^2$. 

$\blacksquare$ \textbf{Weight Fusion:} 
To this end, as shown in the blue box on the left of Fig.~1, a mixing coefficient $\delta_t$ is introduced to fuse the two weight components, yielding the final decision weight $w_t = \text{Normalize}((1-\delta_t) w_k^{ep} + \delta_t w_t^{st})$, where $\delta_t = \text{clip}(\delta_0 + \beta d_t, \delta_{\min}, \delta_{\max})$. Here, $\delta_0$, $\beta$, $\delta_{\min}$, and $\delta_{\max}$ are constant hyperparameters, and $d_t = |(1 - \mathcal{C}(t)) - (1 - \mathcal{R}(t))|$ quantifies the imbalance between the progress of the two subtasks. A larger imbalance leads to a higher value of $\delta_t$, thereby increasing the influence of the step-level weight and allowing more flexible adaptation to the current environment. Given the fused weight vector $w_t = [w_{cov}(t), w_{comm}(t)]^T$, the scalarized Q-value used for action selection is obtained via an inner product between $w_t$  and the task-specific Q-values. Finally, the UAV selects its action according to an $\varepsilon$-greedy strategy , which balances exploration and exploitation, and executes the resulting optimal action $a_t^*$.

\section{Numerical Results}
\subsection{Simulation and Parameter Configuration}
The proposed method is evaluated via PyTorch simulations. 
The mission horizon consists of \(T\) discrete time slots. In the default training setup, we deploy \(M=6\) UAVs and \(N=50\) users over a \(1\,\mathrm{km}\times 1\,\mathrm{km}\) area, discretized into square cells of side length \(L=100\,\mathrm{m}\) (i.e., a \(10\times 10\) grid). UAVs start from the initial point with full battery, fly at constant speed \(v=20\,\mathrm{m/s}\), and maintain a fixed altitude \(h=50\,\mathrm{m}\). We assume no collisions with static obstacles (e.g., buildings). An episode terminates when (i) full-area coverage and complete user-data upload are both achieved, or (ii) all UAVs exhaust their onboard energy. User mobility follows a Gauss--Markov process with mean speed \(\bar v=0.6\,\mathrm{m/s}\), mean turning angle \(\bar\theta=\pi/2\), and memory factor \(\alpha=0.9\). The slot decomposition is \(t_f=5\,\mathrm{s}\), \(t_d=0.01\,\mathrm{s}\), \(t_{\mathit{cov}}=0.1\,\mathrm{s}\), and \(t_{\mathit{comm}}=0.75\,\mathrm{s}\). The propulsion power is set to \(p_m^{\mathrm{prop}}(t)=497.25\,\mathrm{W}\), and the per-UAV energy budget is \(\bar{p}=250\,\mathrm{kJ}\). The field of view of the UAV camera is \(\varphi_h=\varphi_v=90^\circ\), yielding \(L_h = 2h\tan\left(\frac{\varphi_h}{2}\right)=100\,\mathrm{m}\), \(L_v = 2h\tan\left(\frac{\varphi_v}{2}\right)=100\,\mathrm{m}\).
Total bandwidth \(B=16\,\mathrm{MHz}\), number of subchannels \(N=10\), channel-gain threshold \(\underline{h}=2\times 10^{-5}\), transmit power \(p_{\mathrm{tx}}=0.18\,\mathrm{W}\), path-loss exponent parameter \(K_{ps}=2\), Rician factor \(K_s=1\), \(\alpha_{m,n}=5\times10^{-5}\), and noise power \(\sigma^2=1\times10^{-14}\,\mathrm{W}\). The initial data demand per user is \(D_n(0)=100\,\mathrm{Mbits}\). The completion-ratio thresholds are corrected to \(\rho_c=0.8,\rho_r=0.98\).
The hierarchical weight-mixing coefficient is $\delta_t$ with \(\delta_0=0.45\), \(\beta=0.20\), \(\delta_{\min}=0.15\), and \(\delta_{\max}=0.45\). In all DRL-based methods, action selection follows an $\varepsilon$-greedy policy with a unified linear annealing schedule defined as $\varepsilon_t=1 - \dfrac{t}{7500} (1-0.0025)$. Here, Table.~\ref{tab:main_params} shows the other parameter settings and network configurations.

\begin{figure*}[t]
	\centering
	\includegraphics[scale=0.3]{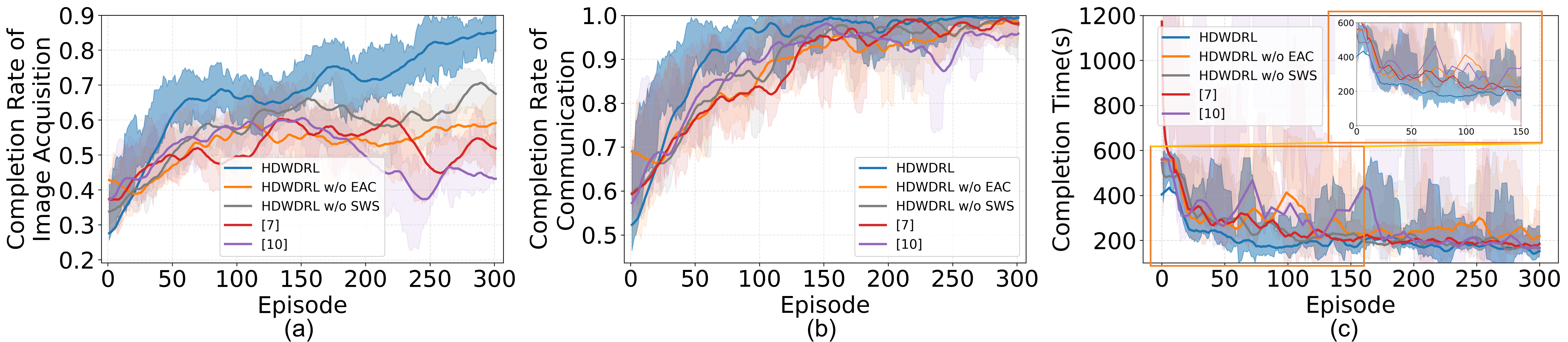}
	\caption{Performance comparison of various methods in terms of (a) image acquisition completion rate, (b) communication completion rate, and (c) completion time.}
	\vspace{-0.6cm}
	\label{fig:2}
\end{figure*}

\begin{table}[htbp]
	\centering
	\caption{MAIN PARAMETERS in HDWDRL FRAMEWORK}
	\label{tab:main_params}
	\begin{tabular}{c|c}
		\hline
		\textbf{Parameters} & \textbf{Values} \\
		\hline\hline
		
		Action dimension (\(|\mathcal{A}|\)) & 4  \\
		\hline
		Episode  & 300 \\
		\hline
		Softmax temperature in step-level weight net (\(\tau\)) & 0.5 \\
		\hline
		Discount factor (\(\gamma\)) & 0.9 \\
		\hline
		EMA smoothing coefficient ($\alpha_{ema}$) & 0.25 \\
		\hline

		Episode weight actor/critic learning rate  & \(1\times10^{-3}\) \\
		\hline
		Step-level weight net learning rate  & \(5\times10^{-4}\) \\
		\hline
		DQN optimizer learning rate & \(1\times10^{-3}\) \\
		\hline
		Replay buffer size  & 8000 \\
		\hline
		Target-network update period & 5 (steps) \\
		\hline
		
	\end{tabular}
	\label{tab:env_params}
\end{table}

\subsection{Performance Evaluation and Analysis}

According to the objective of the system (see (7)),  we conduct extensive experiments to comprehensively evaluate the effectiveness of the proposed  \textbf{HDWDRL} framework. The overall performance of each method is assessed in terms of the completion rate of image acquisition $\mathcal{C}(T)$, communication $\mathcal{R}(T)$ and the task completion time $T$. The results of performance evaluation are shown in Fig.~2, in which 
\textbf{HDWDRL w/o EAC} denotes the proposed HDWDRL framework that removes the episode-level Actor-Critic module (see subsection.~\ref{subsect:3C}) which is used to evaluate the effect on long-term policy stability, and \textbf{HDWDRL w/o SWS} denotes the proposed HDWDRL framework that eliminates the step-wise supervision mechanism (see subsection.~\ref{subsect:3C}) which is used to assess the necessity of short-term weight adaptation. 
Furthermore, both the weight scalarization method \cite{Pan2025IoT} and dynamic weighting method \cite{song2024energy} are also subjected to performance comparisons to verify the performance advantages of HDWDRL framework.

First, we evaluate the overall performance of all schemes in the constructed baseline scenario, as illustrated in Fig.~2(a) and Fig.~2(b). Under the same constraints, the performance of different methods is compared in terms of the completion rate of image acquisition $\mathcal{C}(T)$ and communication $\mathcal{R}(T)$. It can be observed that, after $240$ training episodes, the proposed HDWDRL approach successfully achieves all predefined task thresholds (i.e., $\mathcal{C}(T)\geq 0.8$ and $\mathcal{R}(T)\geq 0.98$) whereas the other baseline methods have not yet reached the requirements. Obviously, this advantage is primarily attributed to the hierarchical dynamic weighting mechanism in HDWDRL, which enables more efficient and stable policy learning in multi-objective settings. Further analysis reveals that the static weighting strategy in \cite{Pan2025IoT} exhibits a relatively slower growth rate in performance, indicating that fixed weights are insufficient to adapt to dynamically changing environments, thereby limiting convergence efficiency. In contrast, although the dynamic weighting method in \cite{song2024energy} introduces a certain degree of adaptability, its performance curve shows significant fluctuations, suggesting that it fails to fully address the instability issue in multi-objective dynamic environments.
Moreover, the ablation results of HDWDRL demonstrate the effectiveness of its key components. Without the episode-level weighting mechanism, the algorithm struggles to maintain long-term policy consistency, while removing the step-level weighting mechanism weakens its ability to respond to instantaneous environmental variations. 

Subsequently, Fig.~2(c) compares the task completion time $T$ required by each method to achieve all task thresholds (i.e., $\mathcal{C}(T)\geq 0.8$ and $\mathcal{R}(T)\geq 0.98$). It is evident that HDWDRL shows less time to complete the tasks and exhibits the lowest performance variance among all methods. At the $80$-th episode of the training process, HDWDRL has almost reached the optimal performance level and gradually converges during the subsequent training process, which demonstrates that the step-level weights enhance short-term responsiveness to dynamic environments, while the episode-level weights ensure long-term stability. Their synergistic effect not only accelerates convergence but also significantly improves the robustness of the learning process.

\section{Conclusion}
This work proposes HDWDRL, a DRL framework for multi-agent collaborative environments. It integrates an episode-level Actor--Critic for global weight adaptation with a step-level lightweight network for short-term weight inference, achieving a convex trade-off between long-term stability and short-term responsiveness. 
Driven by distributed DQN agents, the framework mitigates resource imbalance inherent in static weighting and alleviates policy instability in dynamic weighting, while maintaining a balance between coverage and communication efficiency. Simulation results show that HDWDRL outperforms static and single-scale dynamic baselines in convergence stability, Pareto optimality, and adaptability to dynamic environments.

\bibliographystyle{IEEEtran}

\end{document}